\documentstyle[aps,prl,multicol,epsf]{revtex}
\begin{document}
\newcommand{\tbox}[1]{\mbox{\tiny #1}}
\newcommand{\half}{\mbox{\small $\frac{1}{2}$}}
\newcommand{\sfrac}[1]{\mbox{\small $\frac{1}{#1}$}}
\newcommand{\mbf}[1]{{\mathbf #1}}

\title{Quantum-Mechanical Non-Perturbative Response \\ 
of Driven Chaotic Mesoscopic Systems}

\author
{
Doron Cohen$^1$ and Tsampikos Kottos$^2$\\
\footnotesize
$^1$
Department of Physics, Harvard University, Cambridge, MA 02138 \\
$^2$
Max-Planck-Institut f\"ur Str\"omungsforschung,
37073 G\"ottingen, Germany
}

\date{September 2000}

\maketitle


\begin{abstract}
Consider a time-dependent Hamiltonian ${\cal H}(Q,P;x(t))$ 
with periodic driving $x(t)=A\sin(\Omega t)$.
It is assumed that the classical dynamics is 
chaotic, and that its power-spectrum extends over 
some frequency range $|\omega|<\omega_{\tbox{cl}}$.   
Both classical and quantum-mechanical (QM) linear response 
theory (LRT) predict a relatively large response for 
$\Omega<\omega_{\tbox{cl}}$, and a relatively small response 
otherwise, independently of the driving amplitude $A$. 
We define a non-perturbative regime in the $(\Omega,A)$ space, 
where LRT fails, and demonstrate this failure numerically. 
For $A>A_{\tbox{prt}}$, where $A_{\tbox{prt}}\propto\hbar$, 
the system may have a relatively strong response 
for $\Omega>\omega_{\tbox{cl}}$ due to QM non-perturbative 
effect. The shape of the response function becomes $A$ dependent.  
\end{abstract}

\begin{multicols}{2}

The {\em wall formula} for the calculation of friction in 
nuclear physics \cite{wall}, and the {\em Drude formula} for the 
calculation of conductivity in mesoscopic physics, are just two special 
results of a much more general formulation of `dissipation theory' 
\cite{jar,wilk,vrn,frc}.  
The general formulation of the `dissipation' problem \cite{vrn} 
is as follows: Assume a time-dependent chaotic 
Hamiltonian ${\cal H}(Q,P;x(t))$. For $x=\mbox{const}$ 
the energy is constant of the motion. 
For non-zero $V\equiv\dot{x}$ the energy distribution evolves, 
and the {\em average} energy increases 
with time. This effect is known as {\em dissipation}.  
Ohmic dissipation means that the rate of energy absorption 
('heating') is $d \langle {\cal H} \rangle /dt = \mu V^2$, 
where $\mu$ is defined as the dissipation coefficient.  
In case of periodic driving $x(t)=A\sin(\Omega t)$, 
one should replace $V^2$ by the mean square value $\half (A\Omega)^2$, 
and the dissipation coefficient $\mu(\Omega)$ becomes 
frequency dependent. For simplicity we assume that 
conservative work is not involved in changing $x$.

In case of the wall formula, $(Q,P)$ is a particle moving 
inside a chaotic `cavity', and $x$ controls the deformation 
of the boundary.  Ohmic dissipation (in the sense defined above) 
implies a friction force which is proportional to the velocity, 
where $\mu$ is the `friction coefficient', and $\mu V^2$ is 
the `heating' rate. 
A mesoscopic realization of such system would be a quantum dot 
whose shape is controlled by electric gates.
In case of the mesoscopic Drude formula, $(Q,P)$ is a charged 
particle moving inside a chaotic `ring', and $x$ is the magnetic 
flux through the hole in the ring. 
Ohmic dissipation implies Ohm law, where $V\equiv\dot{x}$ 
is the electro-motive-force, $\mu$ is the conductance, 
and $\mu V^2$ is the `heating' rate. 
For a mesoscopic realization of such system note that 
ring geometry is not important. One may consider a simple 
two dimensional quantum dot driven by a time-dependent 
homogeneous perpendicular magnetic field \cite{wlf}. 
(For the latter geometry it is better not to use the term 
conductance while referring to the dissipation coefficient $\mu$).

In the general analysis of the `dissipation' problem one 
argues that due to the driving there is diffusion in energy space.
This diffusion process is biased because of its $E$ dependence, 
leading to systematic increase of the average energy. This is the 
reason for having dissipation. Therefore we find convenient from now 
on to consider the diffusion coefficient  $D_{\tbox{E}}$ as the 
object of our study. The relation between $d \langle {\cal H} \rangle /dt$ 
and $D_{\tbox{E}}$ constitutes a generalization of the 
so called fluctuation-dissipation relation.

 
\begin{center}
\leavevmode
\epsfysize=0.35in 
\epsffile{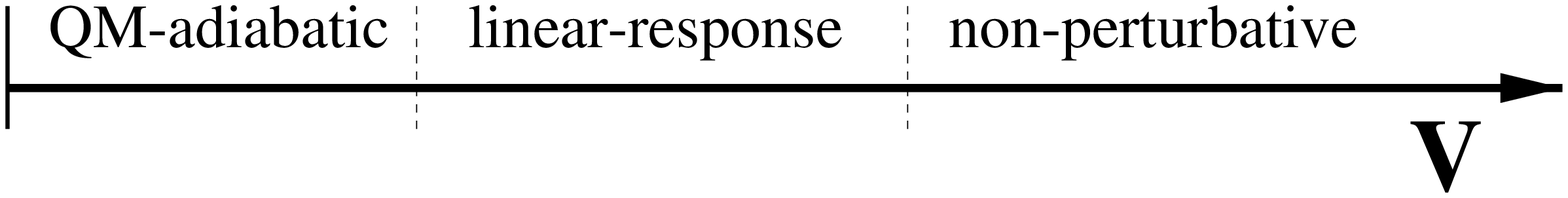}
\end{center}

\vspace*{0cm}

\begin{center}
\leavevmode
\epsfysize=1.8in 
\epsffile{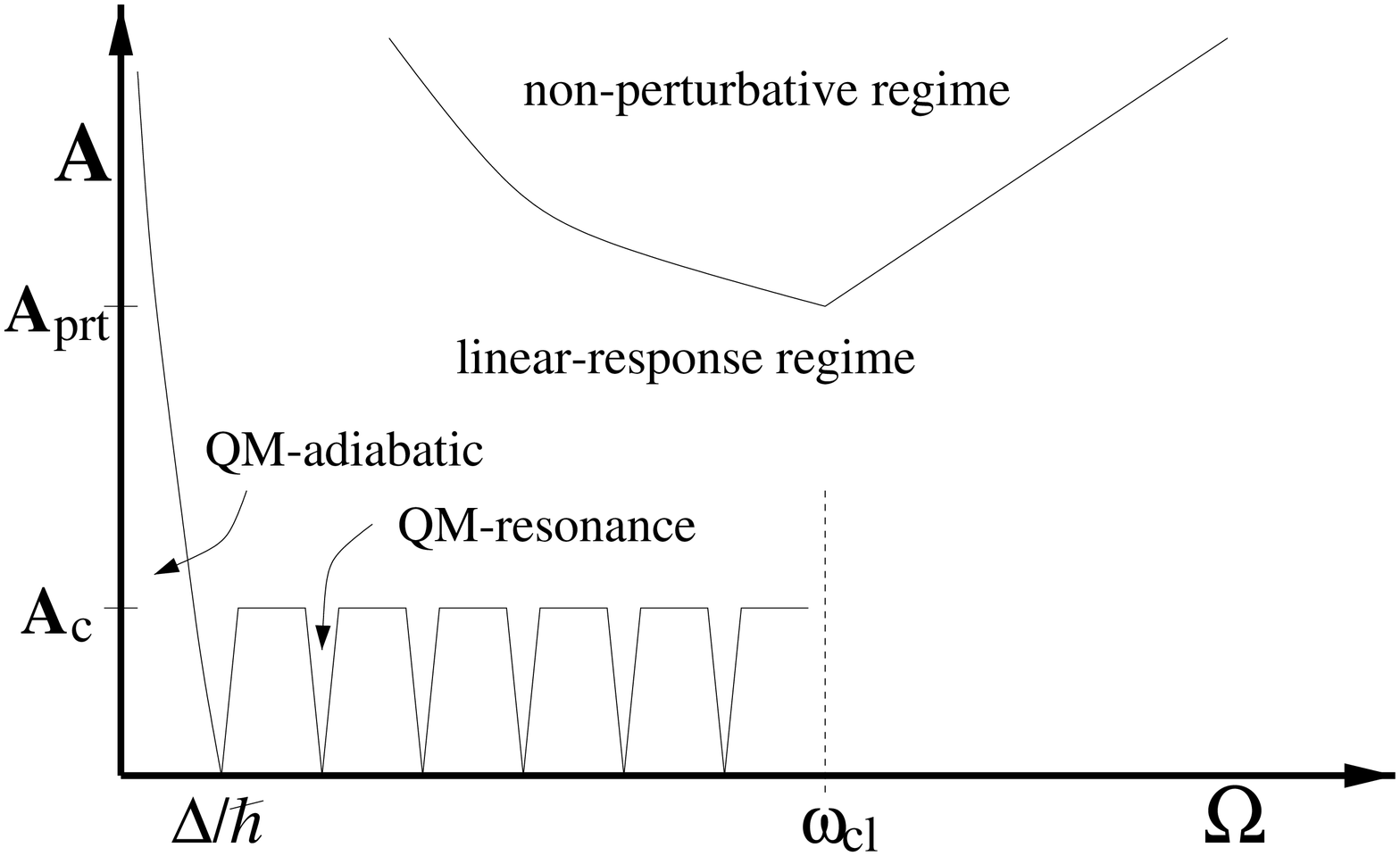}
\end{center}
{\footnotesize {\bf Fig.1.} Upper diagram:
The various $V$ regimes in the theory of quantum dissipation 
for linear driving $x(t)=Vt$. 
Lower diagram: The various $(\Omega,A)$ regimes 
for periodic driving $x(t)=A\sin(\Omega t)$. 
Note the analogy with Fig.5 of Ref.\cite{frc}
with $x \leftrightarrow A$ and $V \leftrightarrow A\Omega$. 
The QM-adiabatic regime (including the regime 
$A<A_c$, but excluding the narrow stripes 
of QM-resonances) is defined by having {\em vanishing} 
first-order probability to go to other levels. 
See the text for further explanations  
and definitions of $A_c$ and $A_{\tbox{prt}}$.}  \\ 
 

Ohmic dissipation is implied if $D_{\tbox{E}}\propto V^2$, 
or $D_{\tbox{E}}\propto A^2$ in case of periodic driving.
Such behavior can be established within the framework of 
classical mechanics \cite{jar,wilk}
using general classical considerations \cite{ott}. 
The classical formulation of the dissipation problem \cite{frc} 
can be regarded as a systematic scheme that justifies 
the use of classical `linear response theory' (LRT).
The precise conditions for the applicability 
of the classical LRT result are further discussed 
in \cite{vrn,frc}, and we are going to mention later on 
what we call `the trivial slowness condition'.   
We are interested in the {\em quantum mechanical} (QM) theory 
of dissipation. The traditional derivation \cite{imry} of 
QM LRT leads formally to the same result as in the classical 
analysis \cite{frc}. Therefore from now on we no longer 
distinguish between the `classical' LRT result and the 
`quantal' LRT result and use just the term `LRT result'. 
As a matter of terminology, it should be noted that the QM formulation 
of LRT, also known as Kubo-Greenwood formalism, is completely 
equivalent \cite{frc} to the well known Fermi golden rule (FGR) picture.

So let us assume that the obvious {\em classical} conditions 
for the validity of the LRT result are satisfied$^{\dag}$. 
Now the question is whether, upon quantization, 
there are {\em additional} \mbox{$\hbar$-dependent} conditions 
for the applicability of the LRT result$^{\dag}$. 
In the traditional quantum mechanical literature, 
as well as in the recent mesoscopic literature, 
the focus is on the consequences of having finite 
mean level spacing $\Delta$. This leads to the identification 
of the QM-adiabatic regime (extremely slow driving),
and to the discussion of either the Landau-Zener mechanism \cite{wilk} 
or else the Debye relaxation absorption mechanism \cite{debye} 
for dissipation, as well as to the discussion of QM-resonances. 
The main observation of \cite{crs} is that there is  
another regime, the non-perturbative regime (see Fig.1),  
where QM LRT is not valid. As strange as it sounds, 
this does not imply a failure of the LRT result. 
On the contrary, another observation of \cite{crs} 
is that the regime where the classical approximation 
applies, is well-contained in the non-perturbative regime, 
hence the LRT result becomes valid again because of 
quantal-classical correspondence (QCC) considerations.   
However, if the system does not have a good classical limit 
(as in RMT models) this `recovery' of the LRT result is not 
guaranteed. Moreover, as we are going to discuss later, 
QCC consideration cannot exclude the possibility of having 
a relatively large quantal non-perturbative response whenever 
the LRT result is small in comparison.

The outline of this letter is as follows: 
{\bf (1)} We extend the theoretical considerations 
of \cite{crs} to the case of periodic driving. 
{\bf (2)} We give a specific example where the LRT 
result fails because of a quantal non-perturbative effect.
{\bf (3)} We comment on the issue of localization.  
{\bf (4)} We discuss the role of QCC considerations in the theory. 
Based on the theoretical considerations, the reader should realize 
that the existence of the non-perturbative regime is not related 
to having finite mean level spacing $\Delta$, but rather to having 
finite bandwidth $\Delta_b = \hbar \omega_{\tbox{cl}}$, 
where $\omega_{\tbox{cl}}$ is the dropoff frequency of 
the LRT response. In the context of mesoscopic physics this 
bandwidth is known as the Thouless energy.

Given ${\cal H}(Q,P;x)$ with $x=\mbox{const}$, 
we can define a fluctuating quantity 
${\cal F}(t)=-{\partial {\cal H}}/{\partial x}$. 
The autocorrelation function of ${\cal F}(t)$ will be 
denoted by $C(\tau)$. The power spectrum $\tilde{C}(\omega)$
is defined as its Fourier transform. 
The intensity of fluctuations is defined as 
$\nu=\tilde{C}(0)$, and it is convenient to define 
the correlation time as $\tau_{\tbox{cl}}=\tilde{C}(0)/C(0)$. 
We assume for simplicity of presentation that 
the single time scale $\tau_{\tbox{cl}}$ completely characterizes 
the chaotic dynamics of the system:   
The power spectrum of the chaotic motion 
is assumed to be continuous, and it is non-vanishing 
up to the cutoff frequency $\omega_{\tbox{cl}}=2\pi/\tau_{\tbox{cl}}$.  
We assume that $\tilde{C}(\omega)$ is vanishingly small 
for $\omega>\omega_{\tbox{cl}}$.

\end{multicols}

\begin{center}
\leavevmode
\ 
\epsfysize=2.2in 
\epsffile{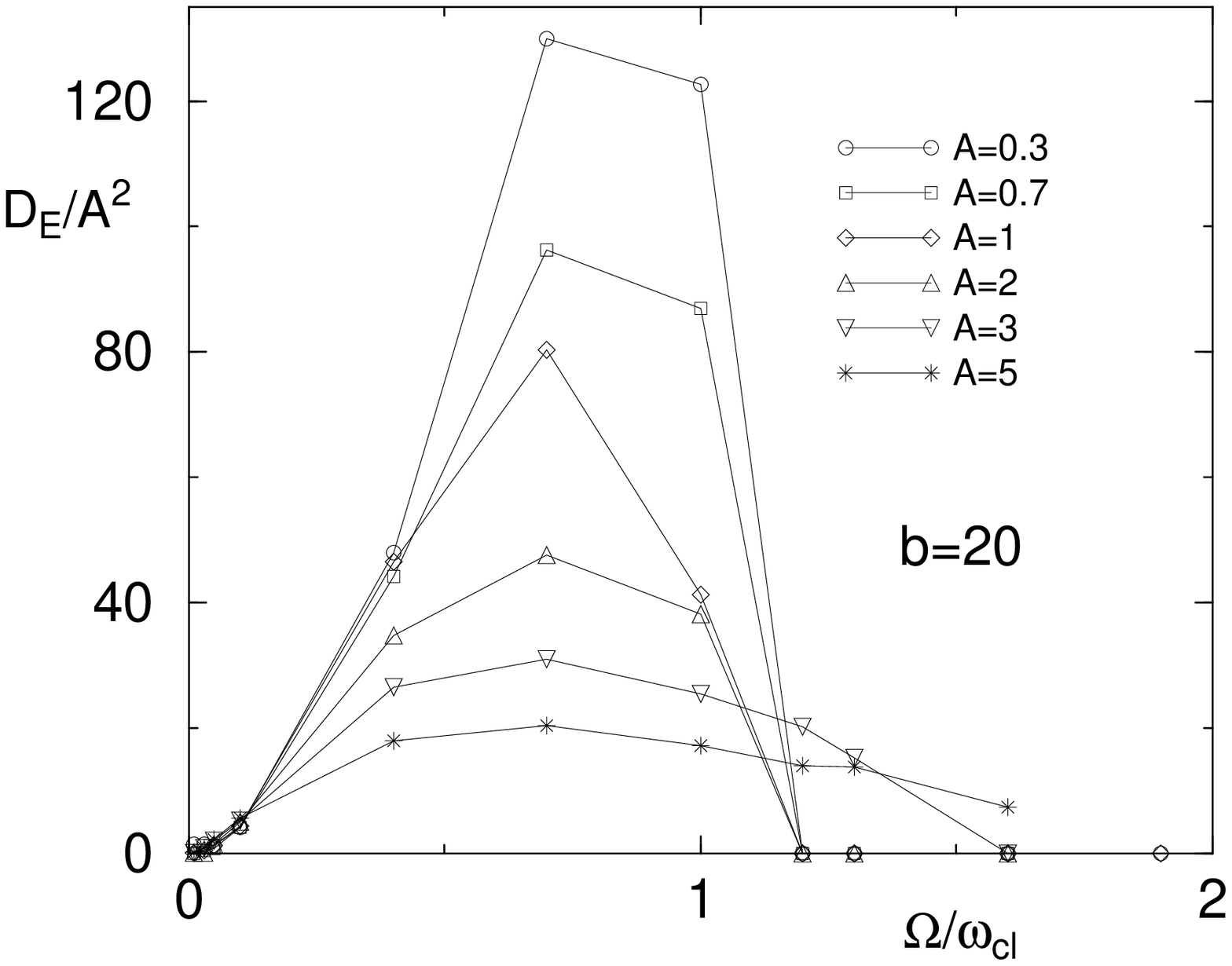}
\hspace*{1.5cm} 
\epsfysize=2.2in 
\epsffile{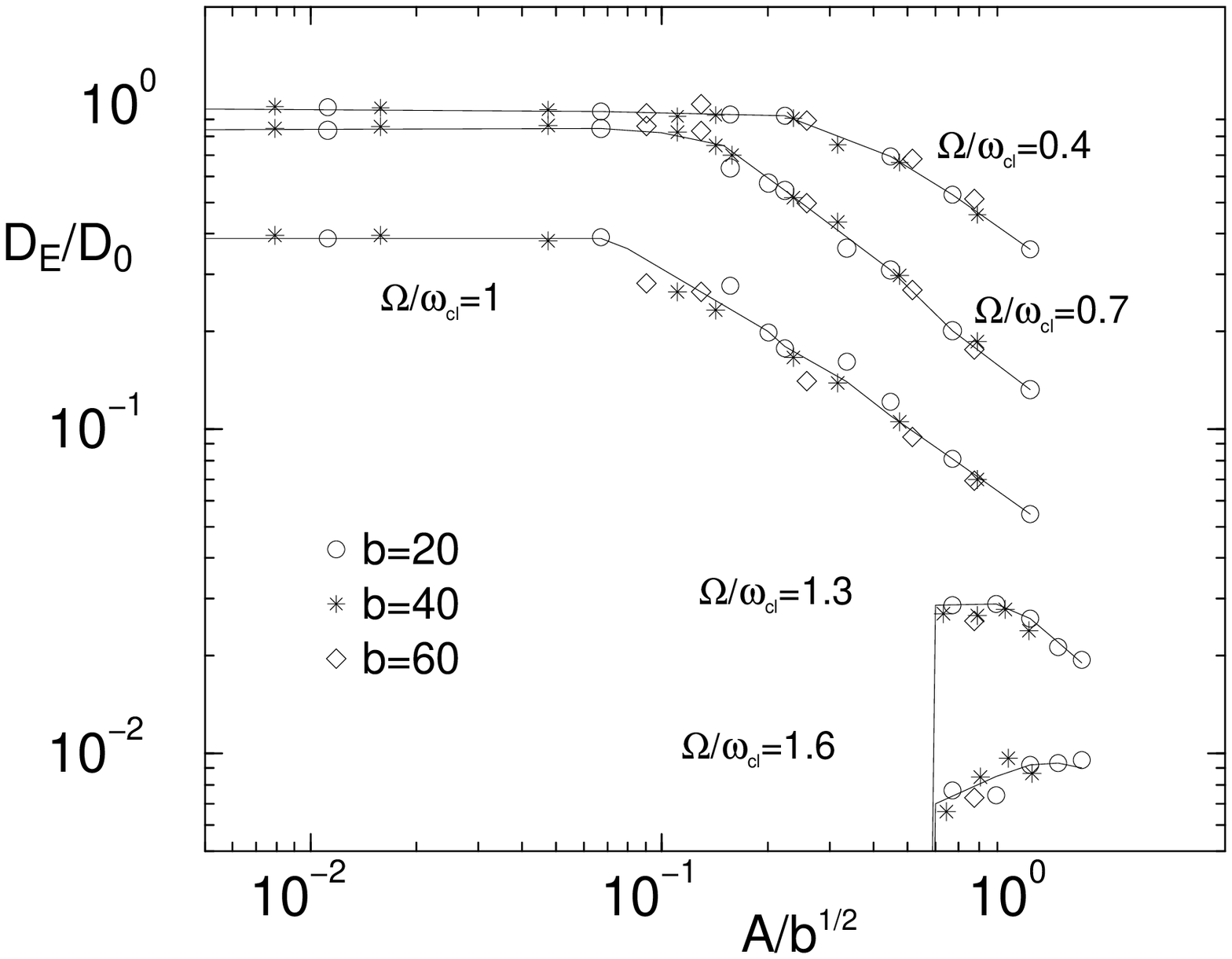}
\end{center}
{\footnotesize {\bf Fig.2.}
The response of a quantum mechanical system is 
displayed as a function of $A$ and $\Omega$. 
The evolution is determined by the WBRM model Eq.(\ref{e5}).
The units of energy and time and amplitude are chosen such that 
$\Delta=0.5$ and $\hbar=1$ and $\sigma=1$ respectively.
{\bf Left:} 
Plots of $D_{\tbox{E}}/A^2$ versus $\Omega/\omega_{\tbox{cl}}$ 
for few values of $A$. For small $\omega$ the plots 
coincide as expected from Eq.(\ref{e1}). 
As $A$ becomes larger the deviations from Eq.(\ref{e1}) 
become more pronounced, and we get response also 
for $\Omega>\omega_{\tbox{cl}}$. 
{\bf Right:}
Plots of $D_{\tbox{E}}/D_0$ versus $A/\sqrt{b}$ 
for few values of $\Omega/\omega_{\tbox{cl}}$. 
The LRT result Eq.(\ref{e1}) implies   
$D_{\tbox{E}}/D_0=1$ for $\Omega/\omega_{\tbox{cl}}<1$ 
and $D_{\tbox{E}}/D_0=0$ for $\Omega/\omega_{\tbox{cl}}>1$.    
The purpose of the horizontal scaling is to demonstrate 
that $A_{\tbox{prt}}$ rather than $A_c$ is responsible for 
the deviation from this LRT expectation.  
Each `point' in the above plots is determined 
by a simulation that involves typically $35$ realizations 
of the evolution until we start to see saturation 
due to dynamical localization effect. 
The typical time step is $dt=10^{-4}$. In each 
step we verify that the normalization is preserved 
to an accuracy of $0.01\%$.  
In order to eliminate finite size effects we 
have used a self-expanding algorithm. 
Namely, additional $10b$ sites are added to each edge 
whenever the probability in the edge sites exceeds $10^{-15}$.
The diffusion coefficient is determined from 
the fitting $\delta E(t)^2 = \mbox{const} \times t^{\beta}$. 
For sub-diffusive behavior ($\beta < 0.86$) 
we set $D_{\tbox{E}}=0$. The preparation of each 'point' 
in the above plots requires $\sim4$ CPU days on Alpha XP1000 machine.  
}  \\ 
 
\begin{multicols}{2}

Consider the time dependent case $x(t)=A\sin(\Omega t)$. 
The LRT result for the diffusion in energy is  
\begin{eqnarray} \label{e1}
D_{\tbox{E}} \ \ = \ \ 
\frac{1}{2} \tilde{C}(\Omega) \times \half (A\Omega)^2
\end{eqnarray}
and we shall use the notation 
$D_0=\mbox{\small $\frac{1}{4}$}\nu(A\Omega)^2$.
The most transparent QM derivation of this result 
is based on the FGR picture. The energy levels of the 
systems are $E_n$, and the mean level spacing is $\Delta$. 
The Heisenberg time is $t_{\tbox{H}}=2\pi\hbar/\Delta$.
The transitions  between levels are determined by the coupling 
matrix elements $({\partial {\cal H}}/{\partial x})_{nm}$. 
It is well known \cite{mario} that for reasonably small $\hbar$ 
this matrix is a banded matrix. 
The bandwidth is $\Delta_b=2\pi\hbar/\tau_{\tbox{cl}}$, 
and the variance of the in-band elements is 
$\sigma^2 = \nu / t_{\tbox{H}}$. It is common to define 
the QM system using 
the four parameters $(\Delta,b,\sigma,\hbar)$ 
where $b=\Delta_b/\Delta$. It is also 
useful to regard the semiclassical relations 
$\tau_{\tbox{cl}}=2\pi\hbar/(b\Delta)$ and 
$\nu = (2\pi\hbar/\Delta)\sigma^2$ as 
definitions, whenever the classical limit 
is not explicitly specified 
[as in Random Matrix Theory (RMT) models].   
The FGR picture implies strong response if and only if 
$\hbar\Omega < \Delta_b$, leading to 
$\Omega<\omega_{\tbox{cl}}$ as in the classical case. 
Using the FGR picture it is straightforward 
to recover $D_0=(\pi/2)(\hbar/\Delta)(\sigma A\Omega)^2$
in agreement with Eq.(\ref{e1}).

The {\em trivial slowness condition} for the applicability 
of {\em classical} LRT \cite{vrn,frc} is 
$V\tau_{\tbox{cl}} \ll \delta x_c^{\tbox{cl}}$. 
Here $\delta x_c^{\tbox{cl}}$ is the parametric change 
that leads to the breakdown of the  
linearization of ${\cal H}(Q,P;x+\delta x)$ with 
respect to $\delta x$. 
Upon quantization there are two other 
parametric scales that become important \cite{frc},  
namely $\delta x_c^{\tbox{qm}}$ and $\delta x_{\tbox{prt}}$. 
The former is the parametric change which is required in 
order to mix neighboring levels, while the latter 
is the parametric change required in order 
to mix all the levels within the band. Hence we define
\begin{eqnarray} \label{e2}
A_c \ \ \equiv \ \delta x_c^{\tbox{qm}} \ = \ 
&  \frac{\Delta}{\sigma}  
& \ \ \propto \ \ \hbar^{(1{+}d)/2}
\\
\label{e3}
A_{\tbox{prt}} \ \equiv \ \delta x_{\tbox{prt}} \ = \ 
& \sqrt{b} \frac{\Delta}{\sigma} 
& \ \ = \ \ \frac{2\pi\hbar}{\sqrt{\nu \tau_{\tbox{cl}}}} 
\end{eqnarray}
The above parametric scales, of 
time-independent first-order perturbation theory (FOPT), 
manifest themselves also in the time-dependent analysis.
FOPT gives the following result for the probability 
to make a transition from an initial level $m$ to some other level $n$, 
\begin{eqnarray} \nonumber
P_t(n|m) = 
\left| \frac{1}{\hbar}
\left( \frac{\partial {\cal H}}{\partial x}\right)_{\!\!nm}
\int_0^t x(t') 
\exp\left(i\frac{(E_n{-}E_m)t'}{\hbar}\right) 
dt' \right|^2
\end{eqnarray}
The total transition probability is $p(t)=\sum_{n}'P_t(n|m)$ 
where the prime imply omission of the \mbox{$n=m$} term. 
In the regime $\Omega<\omega_{\tbox{cl}}$ one obtains
\begin{eqnarray} \nonumber
p(t)=\frac{1}{\hbar^2}\nu A^2 \times
\left\{ \matrix{
(\Omega^2 / \tau_{\tbox{cl}}) \ \sfrac{4} t^4
& \mbox{for} & 0 < t \ll \tau_{\tbox{cl}} \cr
\Omega^2 \ \sfrac{3} t^3  
& \mbox{for} & \tau_{\tbox{cl}} \ll t \ll 1/\Omega  \cr 
\sfrac{2} t & \mbox{for} & 1/\Omega \ll t \ll t_{\tbox{H}} } \right.
\end{eqnarray}
while in the regime $\Omega>\omega_{\tbox{cl}}$ one obtains
\begin{eqnarray} \nonumber
p(t)=\frac{1}{\hbar^2} \frac{\nu}{\tau_{\tbox{cl}}}  
\frac{A^2}{\Omega^2} \times
\left\{ \matrix{
(1{-}\cos(\Omega t))^2 
& \mbox{for} & 0 < t \ll \tau_{\tbox{cl}} \cr
3/2 
& \mbox{for} &  \tau_{\tbox{cl}} \ll t \ll t_{\tbox{H}} } \right.
\end{eqnarray}
In both cases for $t > t_{\tbox{H}}$ we have recurrences, 
and therefore $p(t) \le p(t_{\tbox{H}})$.
[One should be more careful near resonances: 
There  $t_{\tbox{H}}$ should be replaced by $2\pi\hbar/\delta$, 
where $\delta$ is the detuning]. 
The necessary condition for applicability of FOPT 
at time $t$ is that $p(t')\ll 1$ for any $t'<t$, 
which can be written as $p([0,t])\ll 1$. 
The necessary condition for the applicability 
of the FGR picture is $p([0,\tau_{\tbox{cl}}]) \ll 1$. 
This is the FGR condition \cite{frc} which 
guarantees the separation of time scales 
$\tau_{\tbox{cl}} \ll \tau_{\tbox{prt}}$.
The FOPT breaktime $\tau_{\tbox{prt}}$ is defined as the maximal $t$ 
for which $p([0,t]) < 1$.  Now we can define 
a {\em non-perturbative regime} by the 
requirement $p([0,\tau_{\tbox{cl}}]) > 1$. It is 
straightforward to observe that the non-perturbative regime 
is contained in the region $A>A_{\tbox{prt}}$ where 
\begin{eqnarray} \label{e4}
\left(\frac{A_{\tbox{prt}}}{A}\right)\omega_{\tbox{cl}} 
\ < \ \Omega \ < \ 
\left(\frac{A}{A_{\tbox{prt}}}\right)\omega_{\tbox{cl}} 
\end{eqnarray}
The location of the non-perturbative regime is 
illustrated in Fig.1. For completeness of 
presentation we have also indicated the 
subregion in $\Omega<\omega_{\tbox{cl}}$ 
where we have first-order response 
equal to zero. The condition is 
$p([0,\infty]) \ll 1$ or equivalently $p(t_{\tbox{H}}) \ll 1$. 
This region contains the QM adiabatic regime, 
including the region $A<A_c$, but excluding 
the narrow stripes of resonances. 


We have defined the location of the non-perturbative regime, 
but we did not yet give a suggestion 
how Eq.(\ref{e1}) should be modified. 
Using RMT assumptions with regard to 
$({\partial {\cal H}}/{\partial x})_{nm}$, 
and inspired by related studies 
of wavepacket dynamics \cite{wbr}, we expect   
the result
\begin{eqnarray} \label{e6}
D_{\tbox{E}} = (C / \sqrt{v_{\tbox{PR}}}) \times D_0 
\end{eqnarray}
where $v_{\tbox{PR}} = V/(\delta x_{\tbox{prt}}/\tau_{\tbox{cl}})$,  
and $C$ is a numerical constant. A detailed derivation of (\ref{e6}) 
will be presented in the future. (The crucial step is to argue 
that at $\tau_{\tbox{prt}}$ there is a crossover from ballistic 
behavior to diffusion in the sense of \cite{frc}).
For periodic driving this result should be averaged 
over a period leading to 
$D_{\tbox{E}} \propto A^{2{-}\alpha}$ with $\alpha=1/2$.


We wanted to give a numerical example that demonstrate
the non-perturbative response effect.  
Evidently, the simplest is to consider 
a time-dependent version of Wigner's banded 
random matrix (WBRM) model, 
\begin{eqnarray} \label{e5}
{\cal H} \ \  = \ \ \mbf{E}_0 \ + \ x(t) \ \mbf{B}
\end{eqnarray}
where $\mbf{E}_0$ is an ordered diagonal matrix, 
and  $\mbf{B}$ is a banded matrix. This model 
\cite{wigner,casati} is characterized by 
the parameters $(\Delta,b,\sigma,\hbar)$ 
which we have defined previously. For the numerical 
experiment we have assumed rectangular band profile 
such that all the elements $0 < |n-m| \le b$ are taken 
from the {\em same} distribution, and outside 
the band all the elements are identically zero. 
The results of the simulations are summarized in Fig.2.      
It should be realized that WBRM model Eq.(\ref{e5}) 
has a big disadvantage. Namely, unlike the physical 
examples of the introduction, the statistical properties of the 
model are not invariant for $x(t) \mapsto x(t) + \mbox{const}$. 
One may wonder why we do not use one of the two other 
popular variations of Wigner model \cite{WA,rmt},   
eg ${\cal H}=\mbf{E}+(\cos{x})\mbf{B}_1+(\sin{x})\mbf{B}_2$. 
The problem is that for these models one obtains 
$\delta x_{\tbox{prt}} \sim  \delta x_c^{\tbox{cl}} \sim 2\pi$. 
Therefore there is no regime there where LRT fails 
because of {\em quantal} non-perturbative effect$^{\dag}$.
(Such failure requires the generic separation of scales 
$\delta x_{\tbox{prt}} \ll  \delta x_c^{\tbox{cl}}$). 
Thus it seems that the only way to make a RMT model 
$x$-invariant, is to keep it `perturbative' in nature.
The lack of $x$-invariance in the standard WBRM model 
complicates the calculation of the period-averaged $D_{\tbox{E}}$  
and leads to $D_{\tbox{E}} \propto A^{2{-}\alpha}$ 
with $\alpha$ changing gradually from $1/2$ to $1$. 
The numerical analysis of Fig.2b fits well to $\alpha \sim 3/4$.


 
There are two types of localization effects that we had to consider in our
numerical experiments. The ``WBRM model localization'' can be avoided by
using amplitudes $A<b^{3/2}(\Delta/\sigma)$ in order to guarantee that the
instantaneous eigenstates of Eq.(\ref{e5}) are not localized at any time. 
The ``dynamical localization effect'' on the other hand cannot be avoided. 
It is associated with the periodic nature of the driving. Extending standard
argumentation one observes that the eigenstates of the (one period) Floquet
operator have localization length $\xi\times\Delta$, and that the
associated breaktime is $t^*=\xi\times(2\pi/\Omega)$. The two must be
related by $2D_{\tbox{E}} t^*= (\xi\Delta)^2$ leading 
(in the LRT regime) to the result 
$t^* = 2\pi^2(A/A_c)^2 \times t_{\tbox{H}}$. 
In all our numerical experiments the diffusion has been determined
for times where dynamical localization is not yet apparent. Another
possibility, which we have not used, is to add a small noisy component to
the driving, such as to mimic the typical experimental situation of
having dephasing time much shorter than $t^*$.



It is not obvious that the non-perturbative 
behavior that is implied by RMT assumptions, 
and applies to RMT models, should apply also 
to Hamiltonians that possess a well defined 
classical limit. On the contrary, 
the same considerations as in \cite{wbr} can be 
applied in order to argue that RMT considerations 
are not compatible with the QCC principle. 
Here we are going to explain the main idea, 
and to define our expectations.

Taking $\hbar$ to be very small, it is obvious that 
eventually we shall find ourselves in the 
regime where $A\gg A_{\tbox{prt}}$. Let us consider the 
dynamics during a specified time interval $0<t'<t$. 
The time $t$ is chosen to be much larger than $\tau_{\tbox{cl}}$. 
On the basis of QCC considerations, we should be able 
to make $\hbar$ sufficiently small such that the 
quantum evolution becomes similar to the classical 
evolution up to the time $t$. The classical analysis 
implies that during this time the stochastic behavior 
is established. Therefore having detailed QCC during 
the time $t$ implies that the quantal $D_{\tbox{E}}$  
can be approximated by the classical result. 
This leads to a contradiction with the RMT prediction 
Eq.(\ref{e6}) in the domain $\Omega<\omega_{\tbox{cl}}$, 
but {\em  not} in the domain $\Omega>\omega_{\tbox{cl}}$. 
We are going to further explain this last point.

Denote the energy dispersion by $\delta E_{\tbox{qm}}(t)$, 
and the corresponding classical result by 
$\delta E_{\tbox{cl}}(t)$. For sufficiently small $\hbar$ 
it should be possible to make a leading order approximation 
$\delta E_{\tbox{qm}}(t) \approx \delta E_{\tbox{cl}}(t) + \hbar^{\gamma}g(t)$,
with $\gamma > 0$.
In the $\Omega<\omega_{\tbox{cl}}$ regime the first term in this  
approximation is dominant. On the other hand for 
$\Omega>\omega_{\tbox{cl}}$ the first term gives a vanishingly  
small result for $D_{\tbox{E}}$. Therefore, without any contradiction 
with QCC considerations, the second term becomes important. 
Therefore we may have in principle an enhanced quantal response 
for $\Omega>\omega_{\tbox{cl}}$.

In conclusion, we have defined a non-perturbative 
regime in the $(\Omega,A)$ plane, where LRT cannot 
be trusted. We have  demonstrated an actual failure 
of LRT for a particular (RMT) Hamiltonian.
We believe that for 
generic chaotic systems the RMT mechanism for diffusion 
competes with the classical mechanism. The actual 
response of the system is expected to be determined 
by the predominant mechanism.  The study of this 
conjecture is the theme of our future studies.


We thank the Centro Internacional de Ciencias 
(Cuernavaca Mexico) for their hospitality during the 
Quantum Chaos workshop. DC thanks ITAMP for support.


\end{multicols}

\begin{thebibliography}{99}

\vspace*{-1.5cm}

\bibitem[\dag]{note1} The most serious attempt to 
challenge LRT has been done in \cite{WA}. However, 
in view of a later work \cite{rmt}, and as 
explained in \cite{frc}, the suppression of diffusion 
for large $V$ in those studies follows from the violation 
of the trivial slowness condition 
$V\tau_{\tbox{cl}} \ll \delta x_c^{\tbox{cl}}$.  
 
\bibitem{wall} 
J. Blocki et all. {\em Ann. Phys.} {\bf 113}, 330 (1978).
S.E. Koonin et all. Nuc. Phys. A {\bf 283}, 87 (1977).
S.E. Koonin and J. Randrup, 
Nuc. Phys. A {\bf 289}, 475 (1977). 

\bibitem{jar}
C. Jarzynski, {\em Phys. Rev.} {\bf E 48}, 4340 (1993).
C. Jarzynski, Phys. Rev. Lett. {\bf 74}, 2937 (1995).

\bibitem{wilk}
M. Wilkinson, {\em J. Phys. A} 
{\bf 20}, 2415 (1987);
{\bf 21}, 4021 (1988).


\bibitem{vrn} 
D. Cohen, {\em in} Proceedings of the International 
School of Physics `Enrico Fermi' Course CXLIII 
``New Directions in Quantum Chaos'', 
Edited by G. Casati, I. Guarneri and U. Smilansky, 
IOS Press, Amsterdam, 2000. 

\bibitem{frc}
D. Cohen, Annals of Physics {\bf 283}, 175 (2000).

\bibitem{wlf} 
A. Barnett, D. Cohen, and E.J. Heller, nlin.CD/0006041.

\bibitem{ott} 
E. Ott, Phys. Rev. Lett. {\bf 42}, 1628 (1979). 
R. Brown, E. Ott and C. Grebogi, 
Phys. Rev. Lett, {bf 59}, 1173 (1987); 
J. Stat. Phys. {\bf 49}, 511 (1987).

\bibitem{imry} 
Y. Imry, {\em Introduction to Mesoscopic Physics} 
(Oxford Univ. Press 1997). (Appendix A and further references).

\bibitem{debye} 
R. Landauer and M. Buttiker, Phys. Rev. Lett. {\bf 54}, 2049 (1985).

\bibitem{WA} 
M. Wilkinson and E.J. Austin, 
J. Phys. A {\bf 28}, 2277 (1995).

\bibitem{rmt} 
A. Bulgac, G.D. Dang and D. Kusnezov, 
Phys. Rev. E {\bf 54}, 3468 (1996). 

\bibitem{crs}
D. Cohen, Phys. Rev. Lett. {\bf 82}, 4951 (1999). 

\bibitem{mario}
M. Feingold and A. Peres, Phys. Rev. A {\bf 34} 591, (1986).
M. Feingold, D. Leitner, M. Wilkinson, Phys. Rev. Lett. {\bf 66}, 986 (1991).
 
\bibitem{wigner}
E. Wigner, Ann. Math {\bf 62} 548 (1955); {\bf 65} 203 (1957).

\bibitem{casati}
G. Casati, B.V. Chirikov, I. Guarneri, F.M. Izrailev, 
Phys. Rev. E {\bf 48}, R1613 (1993);
\ Phys. Lett. A {\bf 223}, 430 (1996).

\bibitem{wbr} 
D. Cohen, F.M. Izrailev and T. Kottos,  
Phys. Rev. Lett. {\bf 84} 2052 (2000). 

\end{thebibliography}
\end{document}